\documentclass[a4paper,fleqn,usenatbib]{mnras}

\usepackage{newtxtext,newtxmath}
\usepackage{subfig}
\usepackage[T1]{fontenc}
\usepackage{ae,aecompl}

\usepackage{graphicx}	
\usepackage{amsmath}	
\usepackage{amssymb}	

\title[A super Earth planet in the Taurus]{Discovery of a bright microlensing event with planetary features towards the Taurus region: a super Earth planet}

\author[A.A. Nucita et al.]{
A.A. Nucita,$^{1,2}$\thanks{E-mail: nucita@le.infn.it}
D. Licchelli,$^{3,4}$
F. De Paolis,$^{1,2}$
G. Ingrosso,$^{1,2}$
F. Strafella,$^{1,2}$
 \newauthor
N. Katysheva$^{5}$
and S. Shugarov,$^{5,6}$
\\
$^{1}$ Department of Mathematics and Physics {\it ``E. De Giorgi''} , University of Salento, Via per Arnesano, CP-I93, I-73100, Lecce, Italy\\
$^{2}$  INFN, Sezione di Lecce, Via per Arnesano, CP-193, I-73100, Lecce, Italy\\
$^{3}$  R.P. Feynman Observatory, Gagliano del Capo, Lecce, Italy\\
$^{4}$  CBA, Center for Backyard Astrophysics - Gagliano del Capo, Lecce, Italy\\
$^{5}$  Sternberg Astronomical Institute, Moscow State University, Moscow, 119991 Russia\\
$^{6}$  Astronomical Institute of the Slovak Academy of Sciences, Tatranska Lomnica, 05960 Slovakia\\
}

\date{Accepted XXX. Received YYY; in original form ZZZ}

\pubyear{2017}

\begin{document}
\label{firstpage}
\pagerange{\pageref{firstpage}--\pageref{lastpage}}
\maketitle

\begin{abstract}
{The transient event labeled as TCP J05074264+2447555 recently discovered towards the Taurus region 
was quickly recognized to be an ongoing microlensing event on a source located at distance of only $700-800$ pc from Earth. 
Here, we show that observations with high sampling rate 
close to the time of maximum magnification revealed features that imply the presence of a binary lens system
with very low mass ratio components. We present a complete description of the binary lens system  
which hosts an Earth-like planet with most likely mass of $9.2\pm 6.6$ M$_{\oplus}$. Furthermore, the source estimated 
location and detailed Monte Carlo simulations allowed us to classify the event as due to the closest lens system, being at a distance of $\simeq 380$ pc 
and mass $\simeq 0.25$ M$_{\odot}$.}  
\end{abstract}




\section{Introduction}
Since the discovery in 1995 of the first extra-solar planet around a main sequence star  \citep{mayor1995}, an increasing number of 
new candidates accumulated particularly over the last decade. This was mainly due to the Kepler space mission which, according to the latest 
statistics\footnote{The reader may consult the \url{https://www.nasa.gov/kepler/} site for more details.}, allowed identifying
more than 4000 candidates, with more than 2000 planetary objects confirmed by follow-up observations\footnote{See the continuously updated site 
\url{http://exoplanet.eu}.}.  
Extra-solar planets are commonly found via the transit or radial velocity methods  \citep{irwin2008} but 65 of them have been discovered by using the gravitational 
microlensing technique, i.e. searching for a time-dependent magnification of background sources due to intervening lenses.
For an event caused by a single-mass lens, the brightness variation is characterized by a symmetric (and achromatic) light curve, {the so called Paczy\'nski profile}  \citep{pacz1986}. 
In the case of an intervening binary lens, the resulting light curve may change drastically, depending on the binary and source parameters 
and on the overall geometry of the event \citep{schneider1986}. 
This is also true when the star companion is a planetary object. Hence, possible deviations with respect to the {Paczy\'nski light curve} 
offer the unique possibility to detect planets around faint stars. {
Furthermore, this method is sensitive to planet parameters (as the distance from the host star and the star-to-planet mass ratio) in general not accessible to other techniques
and may allow to detect very far extra-solar planets, even in nearby galaxies \citep{ingrosso2009}.
}

In all cases, the probability to detect an ongoing microlensing event is small as the optical depth of the event 
occurrence, which depends on the lens number density along the line of sight, is $\simeq 10^{-6}-10^{-7}$. Analogously, the event rate, which depends on the 
lens number density, the transverse relative velocity among the observer, the lens and the source \citep{derujula1991,griest1991}, results to be, e.g. towards the 
Galactic bulge, 
$\simeq 10^{-6}-10^{-7}$ events per year and monitored star, when 
modelling (as usual) our galaxy with a triaxial bulge \citep{dwek1995} and a double exponential stellar disk \citep{gilmore1989,bahcall1983}. 
It is then natural that microlensing campaigns select targets
in dense regions as the Galactic bulge, the Large Magellanic Clouds or the Andromeda galaxy \citep{ogle2015,lmc2011,andromeda2007,andromeda2010,andromeda2011,andromeda2014}. 
These expectations drastically drop when considering other directions where the density of both lenses and background 
sources is strongly reduced, thus lowering the chances to detect microlensing events. 

Neverthless, microlensing events may in principle occur in any direction and this is exactly what happened in the case of the transient event 
TCP J05074264+2447555 serendipitously discovered by Kojima on UT 2017-10-25.688 towards
the Taurus region (at {J2000} coordinates RA=$05^h07^m42.64^s$, DEC=$+24^{\circ}47^{'}55.5^{''}$), i.e. towards the galactic anticenter {(l=$178.7552^{\circ}$, b=$-9.3278^{\circ}$)}. 
The transient was soon recognized to be a microlensing event and this makes it exceptional as it occurred 
in a direction of the Galaxy with relatively low number of expected lenses\footnote{To the best of our knowledge, there is only another 
microlensing event detected towards a low stellar density region with a lensed star at adistance of $\simeq 1$ kpc (see \citealt{fukui2007} and references therein).}. Furthermore, we show here that high 
cadence photometry allowed us to clearly distinguish planetary features thus classifying the event as due to a binary system possibly 
hosting an Earth-like planet.

The paper is structured as follows: in Section \ref{observations} we give a short review about all the available data and present our high cadence photometry. In Section 
\ref{method} we give details about the data analysis and in Section \ref{discussion} we address our conclusions.

\section{Observations of a bright microlensing event}
\label{observations}

The transient event TCP J05074264+2447555 was serendipitously discovered on UT 2017-10-25.688 towards
the Taurus region, i.e. a direction of the Galaxy with relatively low number of expected lenses. 
The source is cataloged on USNO-A2 as a star with B and R magnitudes of 14.7 and 13.6, 
respectively, and was spectroscopically recognized as a F5V main sequence star \citep{atel10919}. 

TCP J05074264+2447555 was initially explained as a single-lens microlensing event by Jayasinghe et al. (\citeyear{atel10923}) and by Maehara (\citeyear{atel10919}) 
follow-up spectroscopy and, indeed, the corresponding ASAS-SN light curve was 
well fitted by a Paczy\'nski profile so that it was possible to infer the lens model parameters: the time of closest approach 
$t_0\simeq 58058.80$ days (MJD),  the impact parameter $u_0\simeq 0.091$ (in units of the Einstein angle $\Theta_E$) 
and the Einstein crossing time $t_E\simeq 26.7$ days. 
In addition, no contribution of the lens light to the observed luminosity (the so called blending effect) was reported. 

In the meantime,  follow up observations were triggered in the $X$-ray band with the Swift satellite \citep{atel10921} 
with the aim to characterize the high energy emission (if any) of the intervening lens.  
No $X$-ray source at the target position was found but it was reported a brightening in the UV band.

{Following a VSNET (Variable Star Network\footnote{See \citet{vsnet} and  \url{http://www.kusastro.kyoto-u.ac.jp/vsnet/}}) alert, we started to monitor 
the target source at the R.P. Feynman Observatory (hereafter FO), with two 
telescopes of the Astronomical Institute of Slovak Academy of Sciences in Star\'a Lesn\'a (hereafter SL) and, after the event peak, by
using the telescope of the Crimea station of the Sternberg Astronomical Institute (hereafter CSAI, see \url{http://www.sai.crimea.ua/})}. 

Right away, on the data acquired on the night of 31st October 2017, we distinguished features (a descending trend close to the event peak) 
different from the predictions obtained by using a Paczy\'nski model \citep{atel10923}. In addition, follow up data acquired by T. Vanmunster 
(an expert CBA\footnote{See the Center for Backyard Astrophysics - Belgium -  webpage available at \url{www.cbabelgium.com}} 
and AAVSO\footnote{The American Association of Variable Star Observers webpage is available at \url{https://www.aavso.org/}} observer) 
revealed a subsequent ascending branch in the CV (Clear to V) photometry.

Hence, the detected v-shaped profile pushed us to interpret the data as due to a binary lens event \citep{atelnucita}. 

\subsection{Data reduction}

{The SL-CSAI photometric observations were carried with the {Zeiss 600 mm} (with a CCD-camera FLI ML3041, 2048 pixel $\times$ 2048 pixel, pixel size 15 $\mu$m) 
and 180 mm (equiped with a CCD camera SBIG ST10  MXE, 2184 pixel $\times$ 1472 pixel, and pixel size 6.8$\mu$m) telescopes of the Astronomical Institute of 
Slovak Academy of Sciences (in Star\'a  Lesn\'a, Slovakia) and on the 500 mm AZT-5 telescope operating with the CCD camera Apogee Alta U16M 
(4096 pixel $\times$4096 pixel, and pixel size 9$\mu$m), in the Crimean station of the 
Sternberg Astronomical Institute. 

Standard photometric correction of the CCD frames (debiasing, flat-fielding, darking) were applied to all the CCD frames, and 
then aperture photometry was made by using the MAXIM-DL5 package. The star labeled as GSC1849.1964 was used as a reference star and four local check 
stars were measured for more accuracy. We obtained UBVRcIc magnitudes of GSC1849.1964 on the base of SIMBAD database and according to our measurements:  
U = $10.82$, B = $10.83$, V = $10.404$, Rc = $10.07$, Ic =  $9.87$.}

The R.P. Feynman Observatory (FO) is a small astronomical observatory owned and operated by one of us (DL), located in an urban area in Gagliano del Capo, south Italy, 
and dedicated to photometric and spectroscopic projects, mainly for educational purpose, but also in support of small research programs on asteroids, exoplanets 
and variable stars \citep{dla,dlb}. The main telescope is a 300mm f/5.3 newtonian reflector equipped with an ATIK 460EX CCD camera (2750 pixel $\times$ 2200 pixel, pixel size 4.54$\mu$m, pixel scale=1.18 
arcsec/pixel in binning 2x2),  attached to a filter wheel with a Custom Scientific BVr'i' photometric filters set inside. 
The FO observatory is also included in global networks of facilities cooperating in campaigns of time-series photometry 
of cataclysmic variable stars \citep{patterson}.

{As far as the FO data reduction is concerned, five reference stars (in the same field of view of the target) were used for calibration. 
We performed multi aperture differential photometry by using AstroimageJ \citep{collins} with aperture radii 
depending on the night seeing. 

All the available data were also referred to the heliocentric frame of reference, so that the time of each observation is homogeneously measured in HJD.

Note that, because of the inevitable differences in each observatory instrumental set-up, the data sets might not be properly aligned. As described in the following Section, 
we use the method presented in \citet{shv2014} to inter-calibrate the data sets.}

\begin{figure}
\centering
  \includegraphics[width=\columnwidth]{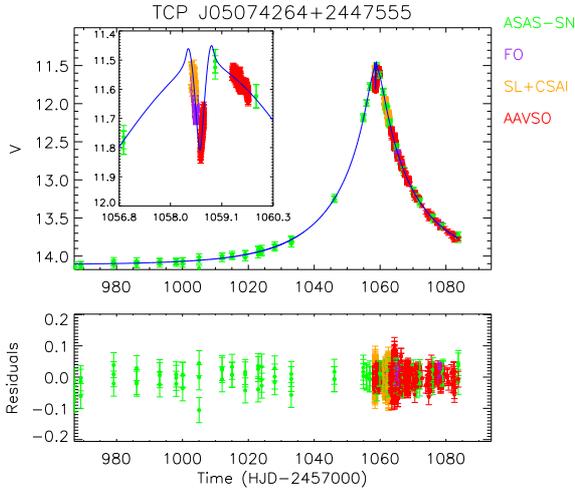}
\caption{The publicly available data from the ASAS-SN (green) and AAVSO (red) databases,  and the data acquired by using the 
FO (purple), SL and CSAI (orange) telescopes are shown on the left panel. {All the data correspond to V band}. 
We also give the best fit curve (blue line) corresponding to the best fit model {\it a}, being the light curve associated to model {\it b} practically indistinguishable.}
\label{figure1}
\end{figure}

{The (V band) inter-calibrated (see next Section) data points from all the available telescopes are in Figure
\ref{figure1} (left panel): SL-CSAI (orange data), FO (purple data), ASAS-SN (green) and AAVSO (red).} 
A close inspection of the data at the time of the event peak allows one to recognize a feature which 
is typical of a binary microlensing event possibly involving a planetary object. 

\begin{figure*}
\centering
\subfloat[]{
  \includegraphics[width=\columnwidth]{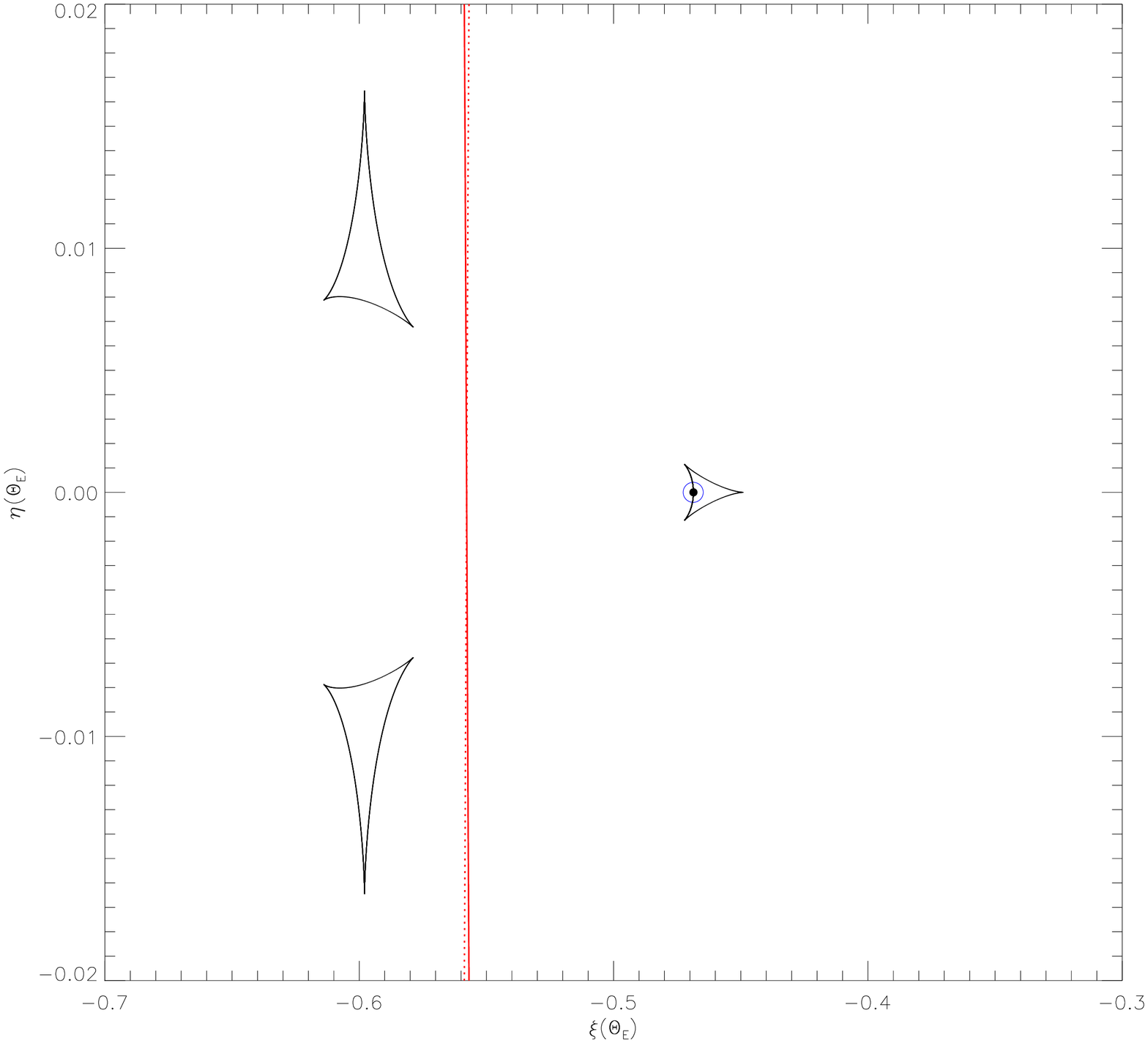}
}
\subfloat[]{
  \includegraphics[width=\columnwidth]{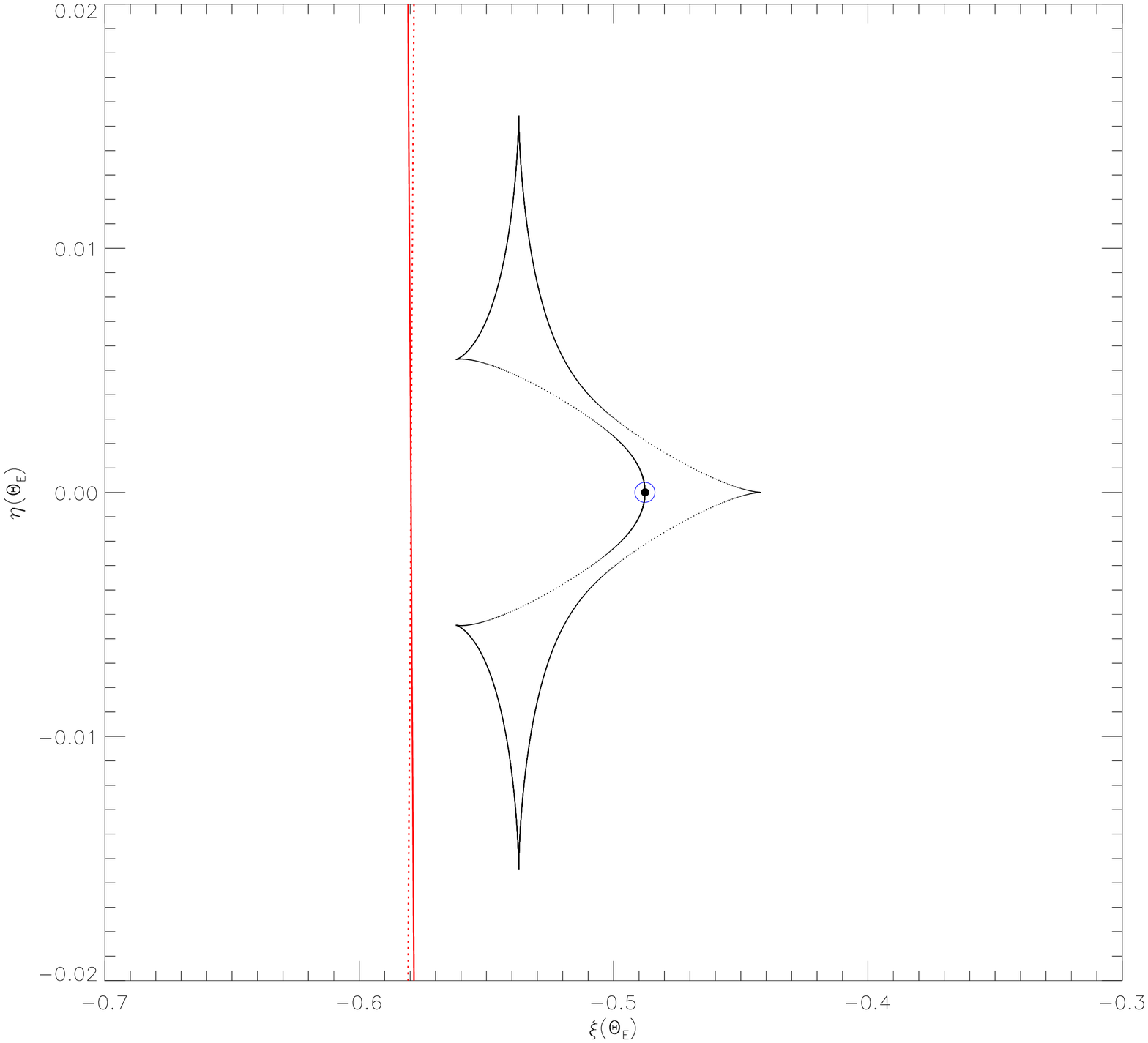}
}
\caption{On the right panel, we show the geometry of the lens system (see text for details) for model {\it a}. The caustics and source trajectory for model {\it b} are shown on left panel.}
\label{figure2}
\end{figure*}

%
%

\section{Modeling the binary lens event}
\label{method}

As shown above, from the data close to the event peak it is clear that a feature which is typical of a binary microlensing 
event is present. We tested this hypothesis and found that by modeling the event with a binary lens magnifying a 
background source it is possible to adequately describe the overall microlensing light curve and reproduce 
the observed features. 

\begin{table*}
\centering
\caption{Best fit results (for models {\it a} and {\it b}) for the TCP J05074264+2447555 light curve. All the errors are quoted at the $68\%$ confidence level (see text for details). 
Angles are in units of the Einstein angular radius. Time is expressed in reduced HJD - $2457000$. }
\label{table1}
\begin{tabular}{|l|l|l|l|}
\hline 
 Parameter                     & {\rm Best Fit }\\
 \hline 
 $q$                           & $(1.1\pm0.1)\times 10^{-4}$               \\
 $b$                           & \begin{tabular}{c}$0.935\pm 0.004$ ({\it a})\\$0.975\pm 0.004$ ({\it b})\end{tabular}          \\
 $T_0$    ({\rm reduced HJD})  & $1058.75\pm 0.01$       \\
 $t_E$    ({\rm days})         & $26.4\pm 0.9$             \\
 $u_0$                         & $-0.093\pm0.001$                        \\
 $\theta$  (rad)               &\begin{tabular}{c}$1.625\pm 0.007$ \\$1.516\pm 0.007$\end{tabular} \\
 $\rho_*$                      & $(6.0\pm0.8)\times 10^{-3}$               \\
 $f_s$                         & $>0.93$                                  \\
 $V_0$                         & $14.11\pm0.01$                            \\
 {\rm Reduced $\chi^2$}/d.o.f  & $1.43/4080$                               \\
  \hline
\end{tabular}
\end{table*}
The best fit procedure consists in adapting a nine parameter model (i.e., the mass ratio $q$, the binary separation $b$, 
the time $t_0$ of the projected closest separation between the source and the binary center of mass, the Einstein time $t_E$, 
the impact parameter $ u_0$ with respect to the center of mass, the source finite size $R_S$, the source trajectory {angle $\theta$} with respect 
to the binary lens axis, the blending factor $f_s$, and the baseline magnitude $V_0$) to the observed data. 
{All angles} are normalized to the Einstein ring angle. For each set of parameters, the associated light curve comes 
from the solution of the lens equation for a generic binary lens system. 
The lens equation in complex notation is given by \citep{witt1990,wittmao1992}
\begin{equation}
\zeta = z+\frac{m_1}{z_1+\bar{z}}+\frac{m_2}{z_2+\bar{z}}
\end{equation}
where $m_1$ and $m_2$ are the masses of the two components (with $m_2<m_1$ so that $q=m_2/m_1<1$), $z_1$ and $z_2$ the positions of the two lenses, 
and $\zeta=\xi+i\eta$ and $z=x+iy$ the positions of the source and image, respectively. 
Here, the adopted reference frame is centred on the mid point of the binary projected axis so that the massive component is on the left while the less 
massive object is on the right. The source position with respect to the binary center of mass (located at distance $-b(1-q)/(1+q)$ from the binary center) 
is given in parametric form by
\begin{equation}
\begin{array}{l}
\xi = u_0\sin(\theta) - \frac{(t-t_0)}{t_E}\cos(\theta)\\
\eta =-u_0\cos(\theta) - \frac{(t-t_0)}{t_E} \sin(\theta)
\end{array}
\end{equation}
where $\theta$ is the angle between the position vector of the massive lens component to the source velocity.
The magnification of each image is obtained by the Jacobian $J$ of the above transformation at the image position, with
the image and source positions corresponding to infinite magnification ($det J = 0$)
forming the critical curves (in the image plane) and  caustic curves (in the
lens plane), respectively. Finally, the total magnification $A(t)$ is obtained summing over all the image contributions.

The solution of the above lens equation can be obtained either
analytically by solving an equivalent 5th order complex polynomial \citep{witt1990,wittmao1992} or 
with the inverse ray-tracing algorithm \citep{schneider1992}. If the former method is fast but unaccurate close to caustics, the latter is robust but very time-consuming 
for best fitting searches unless hybrid solutions are employed \citep{nucita2017}. Alternatively, one can apply {(as in the present work)}
the contour integration scheme via {Green's} theorem \citep{bozza2010} including parabolic corrections in the line integral evaluation. 
This method offers a fast and robust algorithm which allows us to account also for finite source effects. 
In the following we assume for simplicity that the source is extended but with uniform brigthness.

The possible contribution of the primary lens brightness {plus any unresolved stars along the line of sight} to the overall received light, the so called blending effect,
can be accounted for by defining the ratio of the source flux to the total baseline flux, i.e. 
\begin{equation}
f_s = \frac{F_s}{F_b}=\frac{F_s}{F_s+F_{blend}}
\label{blending}
\end{equation}
{so that $f_s=1$ for the case of no blending, while $0<f_s<1$ for blended events.}
Therefore, the final fitting model to the observed data reads out to be
\begin{equation}
m_V=-2.5\log\big[(A(t)-1)f_s+1\big]+V_0 -2.5\log f_s
\end{equation}
where $V_0$ is the baseline magnitude. 

{
In order to find the best fit solution (depending on 9 free parameters) we first sampled the $b$-$q$ parameter space. For each choice of $b$ and $q$ (kept intitially fixed 
at their central values), the other parameters are allowed to vary searching for the minimum value of the $\chi^2$. We then  refined the result 
by associating the contour integration method with a robust minimization algorithm as that offered by the {\it minuit} package \citep{minuit}. 
The errors associated to the $N$ best fit parameters were obtained via a robust  
step procedure consisting in varying the value of one interesting parameter in an adequate range while fixing the values of the remaining $N-1$ parameters 
to the values of the best fit solution. As usual, a variation of $\Delta \chi^2 =1$ for one interesting parameter corresponds to the $68\%$ confidence level. 

Since our data could be affected by systematics due to differente calibrations among the observers and photometry packages, 
we followed the procedure described in \citet{shv2014} (see also \citealt{gould2010}) in order to inter-calibrate the individual data sets. In particular, tha data are aligned to a common microlensing 
magnification model $A(t)$ so that the fluxes observed with the $i$th instrument setup are fit to the linear relationship $F_i(t)=F_{s,i}A(t)+f_{b,i}$, 
where $F_{s,i}$ and $F_{b,i}$ are the instrumental source flux and the blended light coming from the lens and/or any other unresolved source along the line of sight, respectively. 
Note that any difference in the flux zero points is embedded into the $F_{b,i}$ coefficient.

Furthermore, we renormalized the error bars on the data following the procedure described 
in \citet{skowron2011, yee2012}. In particular, once an initial model is found, we rescaled 
the error bars for each separate data set labeled by $i$. Then, for each data set sorted in magnification, 
we evaluated the cumulative distribution of $\chi^2_{i,j}$ as a function of the sorted index $j$, 
renormalized the error bars $\sigma_j$ of the $j$-th point by $K_i\sqrt{\sigma_j^2 +S_i^2}$ and determined the coefficients $K_i$ and $S_i$ 
so that $\chi^2_i/{\rm d.o.f.} \simeq 1$ and the cumulative distribution is a straight line, respectively. After rescaling, we repeated the best fit search and the whole error renormalization algorithm 
described previously.}

\section{Discussion and results}
\label{discussion}

{Following the procedure presented in the previous section, we obtained a binary lens model that adequately describe all the available data whose parameters are reported 
in Table \ref{table1}. Note that in fitting the data to the model, we found two viable 
solutions (with the same $\chi^2=1.43$  for $4080$ d.o.f.) corresponding to $b=0.935\pm0.004$ (model {\it a}) and $b=0.975\pm0.004$ (model {\it b}), respectively. All the other parameters remain
unchanged. It is clear that these solutions, given the associated uncertainties, 
are well separated and then should be considered as a twofold degeneracy.
} 

{In Figure \ref{figure1} we give the publicly available (V band) data from ASAS-SN (green points) and AAVSO (red points) databases, 
and the acquired data by the FO  (purple points), SL and CSAI (orange points) facilities together with the best fit model corresponding to the blue solid 
line. The inset presents a zoomed view around the event peak. In Figure \ref{figure2},  
we show the geometry of the lens system corresponding to the best fit models {\it a} (left panel) and {\it b} (right panel), respectively. The red (solid and dashed) line represents the source (ascending or descending)
trajectory with the red circle and black dot representing the locations of the primary lens object and center of mass, respectively. 
The companion planet (not shown) is located on the positive $\xi$ axis, at symmetric position with respect to the origin (mid point of the 
star-planet separation). The closed paths show the caustic curves. All angles are normalized to the Einstein ring angle $\Theta_E$ corresponding to the 
binary lens total mass. 


Interestingly, the best fit solution corresponds to a mass ratio $q\simeq 10^{-4}$, implying the presence of  a binary lens with a planetary object.} 

{
We remind that the lensed star was recognized to be a $V\simeq 14$ mag star whose spectrum (between 500 nm and 620 nm) is 
similar to that of F5V star HD 31845 \citep{atel10919}. 

An F5V object is a blue to white main  sequence star (with typical mass of $\simeq 1.4$ M$_{\odot}$, radius $\simeq 1.4$ R$_{\odot}$ and $T_{\rm eff}\simeq 6700$ K) 
for which a luminosity of $\simeq 1.48$ L$_{\odot}$ (corresponding to an absolute bolometric magnitude of $3.3$ mag)
is expected. Hence, the distance $D_S$ to the lensed star is estimated to be in the range $700-800$ pc when adopting the full galactic interstellar extinction 
value \cite{extintion} of $A_v\simeq 1.5$) and requiring that the observed $B-V$ color matches that expected for a F5V star.}

{
Therefore, from the estimated source distance $D_S$, the source classification and the best fit source radius $\rho_*\simeq 6\times 10^{-3}$, the Einstein ring angle is found to be
$\Theta_E=R_s/(\rho_* D_s)=1.45\pm 0.25$ mas, where the associated error is obtained by propagating the uncertainties on the relevant parameters. 

We estimate the most probable lens distance $\overline{D_L}$ via a Monte Carlo procedure (see \citealt{yee2012}) and requiring that the events simulated towards 
the TCP J05074264+2447555 direction are characterized by Einstein time and angle values in the respective observed ranges. By modelling our galaxy 
with a triaxial bulge \citep{dwek1995} and a double exponential 
stellar disk \citep{gilmore1989,bahcall1983}, the expected microlensing event rate $\Gamma(D_L, D_S,M_L,v_{\perp}, \cal{M}_S)$ can be evaluated (see, e.g., \citealt{ingrosso2006} 
for details) in any direction, 
being $M_L$, $v_{\perp}$ and $\cal{M}_S$ the lens mass, the observer-lens-source relative transverse velocity and the source absolute magnitude. 

Defining $x=D_L/D_S$ as the dimensionless lens distance, the differential rates $d\Gamma/dx$ and $d\Gamma/dM_L$ can be evaluated after integrating 
over all the remaining relevant quantities.
Therefore, the most probable values for the lens distance and mass are given, respectively, by 
\begin{equation}
\overline{x} = \frac{\int x (d\Gamma/dx) dx}{\int (d\Gamma/dx) dx}, ~~~~~~\overline{M}_L = \frac{\int M_L (d\Gamma/dM_L) dM_L}{\int (d\Gamma/dM_L) dM_L}.
\end{equation}
The Monte Carlo simulation resulted in the distributions in $x$ and $M_L$ (see panels a and b in Figure \ref{probs}) 
characterized by an average 
dimensionless lens distance $\overline{x}=0.51\pm 0.14$ and average lens mass $\overline{M}_L=0.25\pm0.18$ M$_{\odot}$, respectively. Note that the 
lens distance ($\overline{D}_L \simeq 380$ pc) inferred by our Monte Carlo simulation allowed us to classify the transient as due to the closest 
lensing event ever observed.

\begin{figure*}
\centering
\subfloat[]{
  \includegraphics[width=\columnwidth]{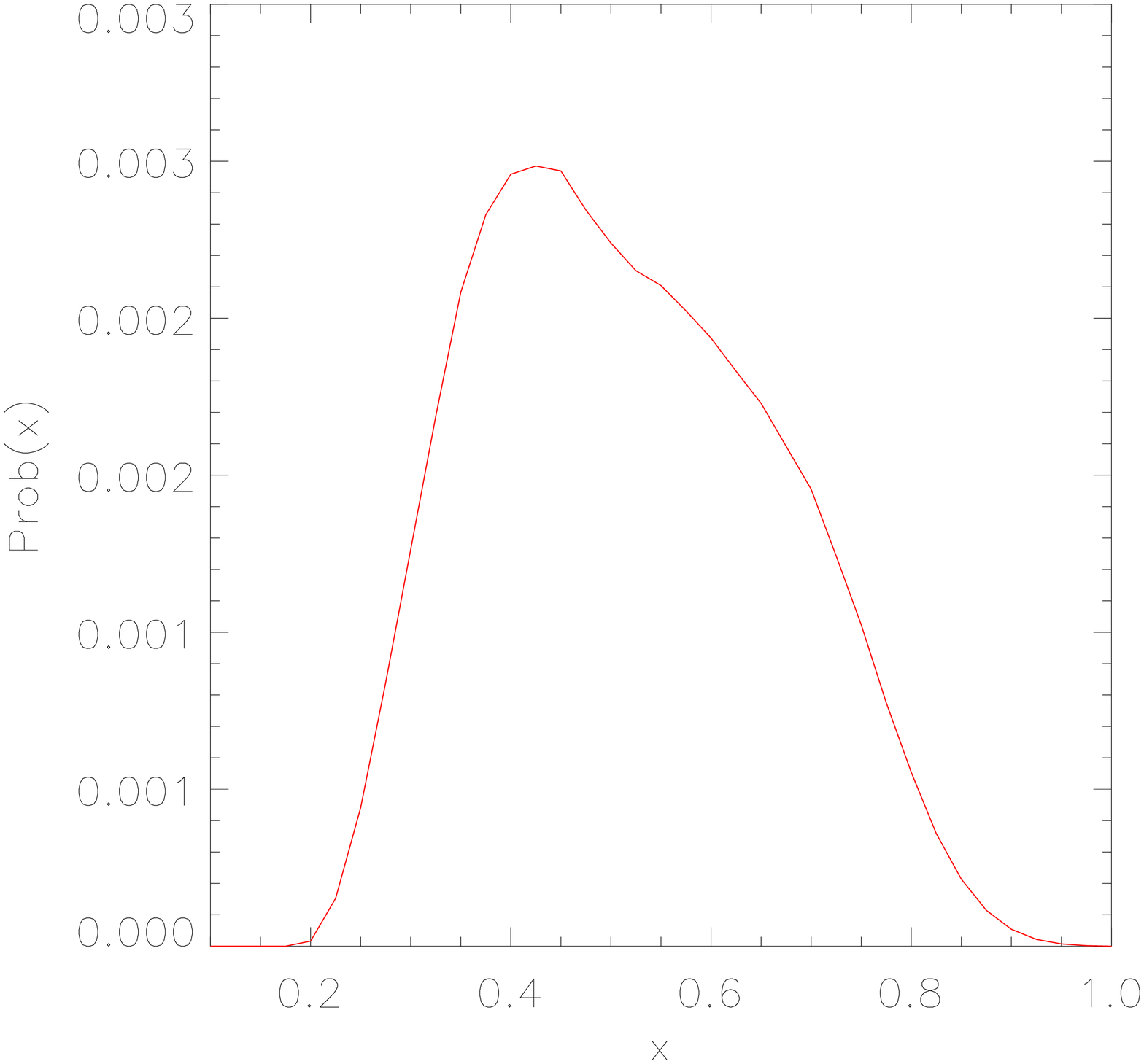}
}
\subfloat[]{
  \includegraphics[width=\columnwidth]{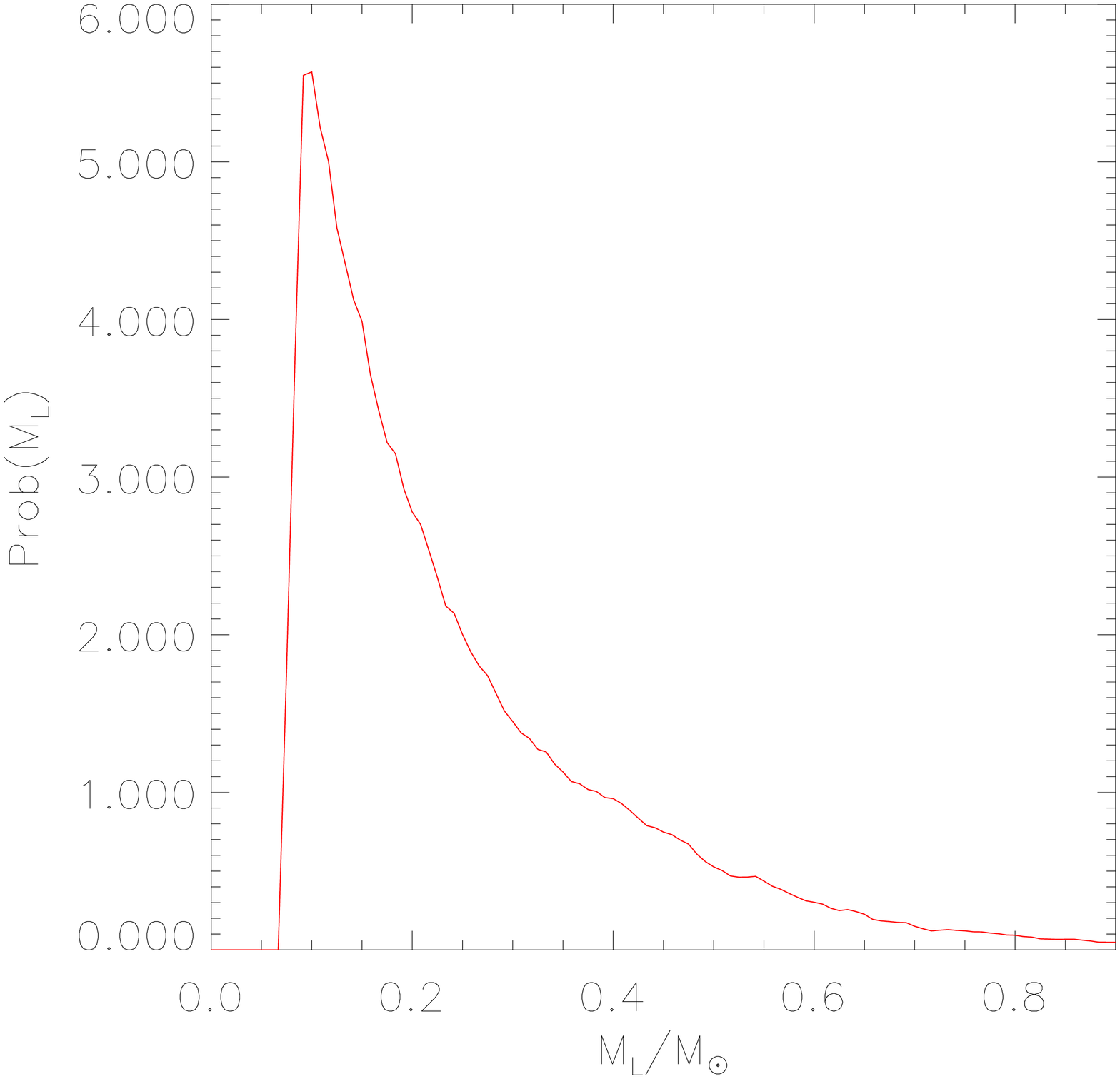}
}
\caption{The probability distributions for the dimensionless lens distance $x$ (left panel) and lens mass $M_L$ (right panel) in units of solar mass (right panel) are shown for the simulated events that 
satisfy the observational constraints (see text for details).}
\label{probs}
\end{figure*}
}

{From the most likely mass of the primary lens object one can infer the mass of the planetary component to be $M_p=9.2\pm 6.6$ M$_{\oplus}$. 
It is then clear that the identified planet (that we nickname here as {\it Feynman-01} in honour of R.P. Feynman) is a planet in the super-Earth 
range orbiting the primary lens object at distance $b\simeq (0.935)$ $\Theta_E$ which is about $0.5$ AU when 
adopting the value of $\overline{D}_L\simeq 380$ pc for the lens distance.


Tighter constraints on the lens system parameters (as planet mass and  the binary semi-major axis) can be obtained by follow-up observations 
allowing to pinpoint the primary lens object. In particular, since the lens is an object with proper motion of $\simeq 20$ mas year$^{-1}$, adaptive optic observations  
could allow to separate it from the source star and eventually confirm the existence of the planetary companion. 
}

\section*{Acknowledgements}
This work makes use of ASAS-SN data (Shappee et al. 2014 and Kochanek et al. 2017). 
We acknowledge with thanks the variable star observations from the AAVSO International Database contributed by observers worldwide and used in this research. 
S. Shugarov thanks for partial support the grants VEGA 2/0008/17 and APPV 15-0458. AAN, FDP, GI and FS thank for partial support the INFN projects TAsP and EUCLID.
{We thank Valerio Bozza and the anonymous Referee for useful suggestions.}






\bsp	
\label{lastpage}
\end{document}